\documentclass[12pt]{article}

\usepackage{
amsmath,
amsfonts,
amssymb,
amscd,
amsthm, 
enumerate, 
}

\usepackage{dsfont} 

\usepackage{geometry}
\usepackage{graphicx}

\newtheorem{theorem}{Theorem}[section]
\newtheorem{proposition}[theorem]{Proposition}
\newtheorem{lemma}[theorem]{Lemma}
\newtheorem{corollary}[theorem]{Corollary}

\theoremstyle{definition}
\newtheorem{definition}[theorem]{\bf Definition}

\renewcommand{\leq}{\leqslant}
\renewcommand{\geq}{\geqslant}

\newcommand{\da}{\downarrow}

\newcommand{\finnat}{\omega^{<\omega}}
\newcommand{\finbin}{2^{<\omega}}
\newcommand{\rar}{\rightarrow}
\newcommand{\computes}{\longrightarrow}
\newcommand{\compeq}{\longleftrightarrow}

\newcommand{\slog}{\text{slog }}
\newcommand{\abs}{\text{abs }}

\newenvironment{acknowledgement}{\noindent\bf Acknowledgment\rm}{}

\newenvironment{keywords}{\noindent\sc Keywords:\rm}{\mbox{}}

\begin{document}

\title{On the equivalence between minimal sufficient statistics, minimal
typical models and initial segments of the Halting sequence}

\author{Bruno Bauwens
\thanks{
Department of Electrical Energy, 
Systems and Automation, Ghent University, 
Technologiepark 913, B-9052, Ghent, Belgium, 
Bruno.Bauwens@ugent.be.
Supported by a Ph.D grant of the Institute for the Promotion of 
Innovation through Science and Technology in Flanders (IWT-Vlaanderen).} 
}

\date{\normalsize \today}

\maketitle  

\begin{abstract}
It is shown that the length of the algorithmic minimal sufficient statistic 
of a binary string $x$,
either in a representation of a finite set, computable semimeasure, or a 
computable function, has a length larger than the
computational depth of $x$, and can solve the Halting 
problem for all programs with length shorter than the $m$-depth of $x$. 
It is also shown that there are strings for which the algorithmic minimal 
sufficient statistics can contain a substantial amount of information
that is not Halting information. 
The weak sufficient statistic is introduced, and it is shown that a 
minimal weak sufficient statistic for $x$ is equivalent to a minimal typical
model of $x$, and to the Halting problem for all strings shorter than the 
$BB$-depth of $x$.
 \\[5pt]
 \begin{keywords}
 $m$-depth -- Kolmogorov complexity -- sufficient statistic --
Halting problem -- typical model -- $BB$-depth
 \end{keywords}
\end{abstract}

\section{Introduction} 

In statistics, a sufficient statistic relative to 
a parametrized family of probability distributions with some 
prior distribution for the parameter, is a function of the data 
that contains enough information to do some Bayesian inference
of the parameter from the distribution that generated the data
\cite{Cover,algorithmicStatistics}.

The definition of an algorithmic sufficient statistic 
 of a string $x$ was introduced as  
the absolute notion of a sufficient statistic from 
statistical theory, it is without reference to a parametrized 
distribution and thus without a prior distribution \cite{algorithmicStatistics}.
The minimal algorithmic sufficient statistic is interpreted as 
``meaningful information" of $x$, \cite{meaningfulInformation, minSuffStatJPG},
and the remaining information of $x$ is interpreted as ``noise". 
This interpretation was applied to image de-noising \cite{minSuffStatJPG}. 
The minimal sufficient statistic can be defined with a representation as
either a finite set, a computable semimeasure, or a computable function.
It is also related to  
the structure function, and therefore to inference methods such as
minimum description length, and Bayesian maximum likelihood 
induction \cite{idealInduction}. It is shown that these induction 
methods perform induction in more or less the same way \cite{structureFunction}.  

In this paper, it is investigated whether ``meaningful information" 
represented by a minimal sufficient statistic
contains the same information as initial segment of the Halting sequence. 
It is shown in proposition \ref{prop:MSSvsMdepth}
that the algorithmic minimal sufficient statistic of a string 
$x$ can compute the Halting problem for all strings with length shorter 
than the $m$-depth.
In Proposition \ref{prop:MSSdiff} it is shown that 
the minimal sufficient statistic can also carry an 
amount of ``noise" or information not related to the Halting problem.
Weak sufficient statistics are introduced, which are within 
super logarithmic\footnote{The super-logarithm is the inverse of the tetration function.
The tetration function is given defined by the sequence: $1, 2,
2^2, 2^{(2^2)}, ...$}
bounds in the length of $x$ also sufficient statistics.
A minimal weak sufficient statistic is constructed and it is shown 
that it is equivalent with some initial segment of the Halting sequence
in Proposition \ref{prop:WSSequivalence}. Finally typical models are investigated and 
the equivalence of a minimal typical model and minimal weak sufficient 
statistic is shown within constant bounds in Proposition \ref{cor:TMvsHalting}.

The minimal sufficient statistic is not computable, and can therefore 
not be directly implemented by any practical computer. However, they can be 
approximated with data-compressors. In this respect many enumerable or 
limit-computable functions in the theory represent
flexible place-holders for programs that can be reused, 
or for programs presenting improving solutions for a task through time.
A theory that tries to interpret algorithms like in
\cite{minSuffStatJPG}, suffices to be accurate within logarithmic bounds. 
However, if one wants to know whether there is a correspondence with 
Halting information, the theory needs to be developed in full detail.

\section{Definitions and notation}

All result here are given in the length conditional setting.
This allows to reduce technical details for some results.
 This choice is also justified by the observation that in most applications of 
statistics or machine learning algorithms the size of the available 
data is predefined or contains no relevant information related to the 
problem.

For excellent introductions to Kolmogorov complexity we refer to 
\cite{GacsNotes, LiVitanyi}.
Let $\omega$ be the set of natural numbers and let for any set $S$, 
$S^{<\omega}, S^n$ be the set of sequences of elements of $S$ of finite length, and of length $n$.
Let $2^{<\omega}, 2^n$ be the sets of the finite binary sequences and binary sequences of length $n$.
The natural association 
$$
 \finbin \rar \omega: \epsilon \rar 0, 0 \rar 1, 1\rar 2, 00 \rar 3, ... 
$$
is implicitly used were needed. For any $x \in \finbin$, $x^i =
x_1...x_i$. 

An interpreter $\Phi$ is a partial computable function:
$$
 \Phi:\omega \times \finbin \times \finnat \rar \finnat: t,p,x \rar \Phi_t(p|x).
$$
and $\Phi(p|x)= \lim_{t \rar \infty} \Phi_t(p|x)$.
The use of $\finnat$ in this definition is to allow $\Phi$ to have
multiple inputs and outputs in $\omega$ or $\finbin$.
An interpreter is prefix-free if for any $x$, the set $D_x$ of all $p$ where 
$\Phi(p|x)$ is defined, is prefix-free.
Let $\Phi$ be some fixed optimal universal prefix-free interpreter. 

For $n \in \omega$, and $x,y \in \finnat$, the \textit{Kolmogorov complexity} $K(x|y)$, is defined as:
\begin{eqnarray*}
 K_t(x|y) & = & \min\{l(p): \Phi_t(p|y,n) \da=x\} \\
 K_t(x) & = & K_t(x|\epsilon).
\end{eqnarray*}
$K(x|y)$ and $K(x)$ are obtained by taking the limit in $t$.
Remark that in the definition of $K$, the parameter $n$ is always
implicitly assumed to be available for $\Phi$.
For functions $f,g$, the notation $f \leq^+ g$ and $f =^+ g$ is used for
 $f \leq g + O(1)$ and $f = g \pm O(1)$.
With abuse of notation let $\log x = \lceil \log x \rceil = l(x)$. Remark that given 
$\log x$, $x$ can be decrypted by its binary representation, and therefore: 
$K(x|\log x) \leq^+ \log x$.

Prefix-free Kolmogorov complexity is additive:
\begin{equation} \label{eq:Kadditivity}
 K(x,y) =^+ K(x) + K(y|x^*),
\end{equation}
where $x^*$ is a program of length $K(x)$ that outputs $x$.

For $x,y \in \finnat$, $n\in \omega$, 
$$
 x \computes y
$$ 
means that there is a program $p_x$ with $l(p_x) \leq O(1)$, 
such that $\Phi(p_x|y,n) \da= x$. Remark that $\Phi$ is also conditioned to $n$. 
Also remark that if $x \computes y$, than $K(x) \geq^+ K(y)$.
\begin{lemma} \label{lem:generateWithnesses}
 For any $w,p \in \omega$ with $\Phi(p) \da= w$ and 
 $l(p) \leq^+ K(w)$ we have 
 \begin{equation} 
  w^* \computes p.
 \end{equation} 
\end{lemma}
\noindent
This is shown in \cite{decomposition}, or
follows from combining the results of \cite[Exercise ...]{LiVitanyi}. 

The complexity of a finite set $S$ is the minimal length of a program on $\Phi$ 
that enumerates all elements of $S$ and halts. The complexity of a computable function 
$f$ is: $\min \{l(p): \forall x[ \Phi(p|x) \da= f(x)]\}.$

A length conditional semimeasure is a positive function $P$ such that for every $n$:
$$
 \sum \{P(x): x \in 2^n\} \leq 1.
$$
From now on, only length conditional semimeasures are used, and referred as semimeasures.
A semimeasure $P$ multiplicatively dominates a semimeasure $Q$, notation $P \geq^* Q$
 if there is a constant $c$ such that $cP \geq Q$. $P =^* Q$ means $P \leq^* Q$ and
$Q \leq^* P$.
A semimeasure $P$ is universal in a set $S$ of semimeasures 
if $P \in S$ and $P$ dominates all semimeasures in $S$.    
Let $m$ be a universal semimeasure 
in the set of enumerable semimeasures.  
By the coding theorem it satisfies: 
$$
 -\log m(x) =^+ K(x).
$$

\section{$m$-depth}

$m$-depth is studied in detail in \cite{mdepth}, where it is defined 
depending on the choice of the universal semimeasure.  
It is also used in \cite{LuminyTalk}, 
where its logarithm appeared as a bound for an on-line coding result. 
$m$-depth can be interpreted as an alternate notion of ``sophistication" of 
a string \cite{sophistication}. 
Here it suffices to use $m$-depth for the specific choice of 
$m_t(x) = \sum \{2^{l(p)}: \Phi(p) \in 2^n\}$. 
This $m$-depth was introduced in \cite{complexityOfComplexity, GacsNotes} without being named. 
It is shown that it dominates Buzzy Beaver depth and 
coarse sophistication \cite{Antunes, mdepth}, however it is more unstable 
in the sense that the $m$-depth of a binary string can vary unboundedly for 
small changes of the constants in its definition \cite{mdepth}.

\begin{definition} \label{def:depth}
The {\em computational $m$-depth} $k_{x}$ of $x \in 2^n$ is given by:
\begin{eqnarray}
 \Omega^n_t &=& \sum \{2^{-l(p)}: \Phi_t(p) \in 2^n\} \nonumber \\
 \Omega_t &=& \lim_t \Omega^n_t\nonumber  \\
 t_k &=&\min\{t: \Omega^n - \Omega^n_t \leq 2^{-k}\} \nonumber  \\
 k_{x} &=& \min\{k:K_{t_k}(x) =^+ K(x)\}. \label{eq:defMdepth}
\end{eqnarray}
Let $\Omega^{n,j}_t$ be the first $j$ bits of the binary expansion of $\Omega^n_t$. 
The {\em Halting sequence $H$}, is given by: 
$$
 H^n_i = \begin{cases} 1 & \text{ if $\Phi(i|n) \da$} \\
		     0 & \text{ otherwise.} \end{cases}
$$
\end{definition}

According to Lemma \ref{lem:complexityOmega}, some initial segment of $H$ and $\Omega^n$ carry 
the same information, and the binary expansion of $\Omega^n$ is incompressible.
\begin{lemma} \label{lem:complexityOmega}
 For $j\leq n$: $\Omega^{n,j} \computes H^{n,2^{j-O(1)}}$ and $K(\Omega^{n,j}|n) \geq^+ j$.
\end{lemma}

The proof is identical as in \cite[Claims 3.6.1, 3.6.2]{LiVitanyi}, and is repeated 
using $m$-depth. 

\begin{proof}
Remark that $\Omega^{n,j} \computes t_j$, by searching for the smallest $t$ with 
$\Omega^{n,j}_t \geq \Omega^{n,j}$.
Any halting program of length shorter than $j - O(1)$,
defines some program that outputs a string in $2^n$, and therefore contributes at least
$2^{-j}$ to $\Omega^{n,j}$.
By definition of $t_j$, any program of length less than $j - O(1)$ that halts, 
must have computation time below $t_j$. Consequently, 
$$
  \Omega^{n,j} \computes t_j,j \computes H^{n,2^{j-O(1)}}.
$$ 

\noindent
If $K(\Omega^{n,j}) \leq j - c$, for $c$ large enough, the corresponding 
program generating $\Omega^{n,j}$, can be turned into a halting program with 
computation time larger than $t_j$, contradicting the previous paragraph.
\end{proof}

Let $K^H(x|y)$ be the Kolmogorov complexity relative to $\Phi^H$, it is $\Phi$ with an 
oracle that contains the Halting sequence $H$ and let $I(x; H) = K^H(x)
- K(x)$. 
\begin{proposition} \label{prop:mDepthVsIH}
 $$
  x,K(x), k_x \computes \Omega^{n,k_x},
 $$
 and
 \begin{eqnarray*} 
  K(x|n) &=^+& k_x + K(x|\Omega^{n,k_x},n) \pm 2\log k_x \\
  I(x;H) &\geq^+& k_x - 2\log k_x.
 \end{eqnarray*} 
\end{proposition}

\begin{proof}
 First an alternate characterization of $m$-depth is given.
 Let $p_1, p_2, ... $ be an enumeration of all Halting programs ordered
by Halting time, it is, for all $t$ if $j < i$ and $\Phi_t(p_j) \da$, 
than $\Phi_t(p_i) \da$. 
Let $p$ be a halting program, and let $i$ such that $p=p_i$, than let
$$
 \alpha_p = \sum \{2^{-l(p_j)}: 1 \leq j \leq i\}. 
$$
Let $\beta_p$ be the first $l(p)$ bits of $\alpha_p$ in its binary
expansion. Remark that the the set of all $\beta_p$ for all 
halting programs $p$ is prefix-free, and 
$$
 p \compeq \beta_p.
$$
Let $\gamma_p$ be the largest prefix $\beta_p^l$ of $\beta_p$
such that $ \Omega^{n} - \beta_p \leq 2^{-l}$. 
Let $x^*$ be the first program in the enumeration $p_1,p_2,...$ 
with $\Phi(x^*) \da=x$ and $l(x^*) =^+ K(x)$ with the same constant as 
implicit in equation \eqref{eq:defMdepth}. 
Let $s$ be the computation time of $x^*$. Remark that $t_{k_x-1} \leq s \leq t_{k_x}$. 
Therefore 
$$
 \Omega^{n} - 2^{-k_x +1} \leq \alpha_{x^*} \leq \Omega^{n} - 2^{-k_x},
$$
and it follows that
\begin{equation} \label{eq:gammaVsOmega}
  \gamma_{x^*} \compeq \Omega^{n,k_x-1}.
\end{equation} 
%
This shows that
$$
 x,K(x),k_x \computes x^*,k_x \computes \alpha_{x^*}, k_x \computes \Omega^{n,k_x},
$$
Which shows the first claim of Proposition \ref{prop:mDepthVsIH}.

Remark that the set of all $p$ such that 
$\gamma_{x^*}$ is a prefix of $\beta_p$ is prefix-free.
Therefore, given $\gamma_{x^*}$, the remaining $l(x^*) - k_x$ bits of 
$\beta_{x^*}$ define a halting program for $x$ given $\gamma_{x^*}$.
Consequently, 
$$
 K(x|\Omega^{n,k_x}) \leq^+ l(x^*) - k_x =^+ K(x) - k_x.
$$
Remark that $K(\Omega^{n,k_x}) \leq^+ k_x + 2\log k_x$. 
Therefore:
\begin{eqnarray*}
 K(x|\Omega^{n,k}) &\geq^+& K(x|(\Omega^{n,k})^*) \\ 
		&=^+& K(x,\Omega^{n,k}) - K(\Omega^{n,k}) \\
		&=^+& K(x,k) - k_x -2\log k_x \\
		&\geq^+& K(x) - k_x  -2\log k_x. 
\end{eqnarray*}
This shows the second claim of Proposition \ref{prop:mDepthVsIH}.

It remains to show the last claim.
Remark that $k,H \computes \Omega^{n,k}$.
\begin{eqnarray*}
 K^H(x) &\leq^+& K^H(x|k_x) + 2 \log k_x \\ 
	&=^+& K^H(x|k_x, \Omega^{n,k})  + 2 \log k_x \\
	&\leq^+& K(x|k_x, \Omega^{n,k})  + 2 \log k_x \\
	&\leq^+& K(x) - k_x  + 2 \log k_x.
\end{eqnarray*}
Therefore, 
$$
 I(x;H) = K(x) - K^H(x) \geq^+ k_x - 2 \log k_x.
$$
\end{proof}

In the proof of Proposition \ref{prop:MSSvsMdepth}, it will be shown that the $\log k_x$-terms 
are necessary. The construction of an explicit weak sufficient statistic in Section \ref{sec:WSS} 
can be considered as an exact variant of this Proposition.

\section{Algorithmic sufficient statistics}

The algorithmic minimal set sufficient statistic was introduced in 
\cite{algorithmicStatistics}. The probabilistic and function variants
are introduced in \cite{meaningfulInformation}.
 For technical reasons the length conditional variants are used here. 
\begin{definition} \label{def:SS}
\begin{itemize} 
 \item A finite set $S$ is a {\em sufficient set statistic} of a binary string $x$ iff $x\in S$ and
\begin{equation} \label{eq:setSS}
   K(S) + \log |S| =^+ K(x).
\end{equation} 
 \item A computable semimeasure $P$ is a {\em sufficient probabilistic statistic} of a binary string $x$ iff 
  $$
   K(P) - \log P(x) =^+ K(x).
  $$
 \item A computable prefix-free\footnote{
	Remark that, it is required here that $F$ is
	prefix-free, as in contrast with \cite{meaningfulInformation}. 
	If $F$ was not required to be prefix-free, than
	it follows that there are strings with $K(F) + l(d) \leq^+ K(x) - \log n$.
  }
  function $F:\omega \rar \omega$ is a {\em sufficient function
statistic} of a binary string $x$  iff for some $d \in F^{-1}(x)$,
  $$ 
   K(F) + l(d) =^+ K(x).  
  $$
\end{itemize}
\noindent
For $Z=S,P,F$, a {\em minimal sufficient statistic} $Z_x$ is the sufficient statistic $Z$ such that $K(Z)$ is minimal
within a constant.
Let $l^Z_x = K(Z_x)$.
\end{definition}

For $Z=S,P,F$, let $|| \log Z ||$ be either: $\log |S|$, $- \log P(x)$ or $\min
\{l(d): d \in F^{-1}(x)\}$. 
The definitions of a sufficient statistic (SS) are summarized by: 
$$
 K(Z) + || \log Z || =^+ K(x).
$$

\begin{proposition} \label{prop:SSmodels}
 Every probabilistic SS of $x$ generates a functional SS of $x$. 
 Every functional SS of $x$ generates a probabilistic SS of $x$. 
\end{proposition}

\begin{proof} The first claim of the proof is solved by 
 applying Shannon Fano coding \cite{LiVitanyi}.
 Suppose $P$ is a SS of $x$, than let for any $y$:
 $$
  \alpha_y = \sum \{P(z): z \leq y\},
 $$
 and let $\beta_y$ be the first $-\log P(y)$ bits of $\alpha_y$.
 Let 
 $$
  F:\omega \rar \omega: \beta_y \rar y.
 $$
 Remark that $F$ is computable, injective and prefix-free.
 If $d$ is the inverse of $x$, than $l(d) = -\log P(x)$, therefore, 
 $F$ is a SS.

 The second claim of the proposition is now shown.
 Suppose $F$ is a SS of $x$, than let for any $y$:
 $$
  P(y) = \begin{cases} \max\{2^{-l(d) - 1}:F(d)=y\} & \text{ if $\exists
							d<2y[F(d)=y]$} \\
			\frac{1}{4y^2} & \text{ otherwise.}
	 \end{cases}
 $$
 Remark that $P$ is computable and that $P$ is a semimeasure:
 \begin{eqnarray*}
  \sum P(y) \leq 1/4 \sum_{i \in \omega} 1/i^2 + 1/2 \sum \{2^{l(d)}: d
\in \text{dom }  F\} \leq 1.
 \end{eqnarray*}
\end{proof}

Remark that $S_x \computes |S_x| \computes \log |S_x|$. Let $S_x^*$ be the shortest program 
that enumerates $S$ and halts, and let $i$ be the index of $x$ in this enumeration. 
A prefix-free encoding of $x$ using $S_x^*,i$ requires $K(S_x) + \log |S_x| + O(1)$. Therefore, 
if $i$ is the index of $x$ in that enumeration, than using Proposition
\ref{prop:mDepthVsIH}:
\begin{equation} \label{eq:minstatComputes}
 S_x^*, i, k_x \computes x, K(x), k_x \computes \Omega^{n,k_x}. 
\end{equation}
where $x^*$ is the witness of $K(x)$ and, $\Omega^{n,k_x}$ 
are the first $k_x$ bits of $\Omega^{n,k_x}$.  
The question rises whether
$$
 S_x^*,k_x \compeq \Omega^{n,k_x},
$$
and if not, how do these differ ? 
An analogue argument holds for the probabilistic and the function case.


\begin{proposition} \label{prop:MSSvsMdepth}
 For all $x$: 
 $$
  \begin{array}{rcl}
    S_x^*, k_x   & \computes &   \Omega^{n}  \\ 
      l^S_x      &   \geq^+  &  k_x - 2\log k_x.
  \end{array}
 $$
\end{proposition}

\begin{proof}
Let $s$ be the computation time of the program $S_x^*$, the shortest program 
of length $K(S_x|n)$ that computes $S$ from $n$. 
Let $f$ be a large enough computable function such that using equation 
\eqref{eq:minstatComputes} it follows that: $K_{f(s)}(x) \leq^+ K(x)$, 
and therefore $s \geq t_{k_x - 1}$. This shows that 
$$
 S_x^*, k_x\computes f(s), k_x \computes t_{k_x-1}, k_x \computes \Omega^{n,k_x-1} \computes \Omega^{n,k_x}.
$$
By Lemma \ref{lem:complexityOmega}, $K(\Omega^{n,k_x}) \geq^+ k_x$ and therefore, 
$l^S_x \geq^+ k_x - 2\log k_x$.
\end{proof}

\begin{lemma} \label{lem:setMinSuffStatVsOther}
 For all $x \in 2^n$ with $K(x) = l(x)/2$:
 \begin{eqnarray*}
  l^P_x &\geq^+& l^S_x \\
  l^F_x &\geq^+& l^S_x.
 \end{eqnarray*}
\end{lemma}

In \cite{meaningfulInformation} it is shown that every set SS
generates a probabilistic SS and every probabilistic SS generates a 
function SS.
Below it will be shown that every function SS generates a set SS if $K(x)$
can be computed from $n$. Since a function SS generates a probabilistic
SS, this finishes the proof.
\begin{proof}
Given $F_x$, the set 
\begin{eqnarray*}
 S^F &=& \{f(y) : y \in 2^{n/2 - l^F_x}\}  
\end{eqnarray*}
contains $x$, and has $K(S^F) \leq^+ l^F_x$. Moreover
\begin{eqnarray*}
 \log |S^F| & \leq^+ & n/2 - l^F_x \leq^+ K(x) - K(S^F).  
\end{eqnarray*}
Therefore, $l^P_x \geq l^S_x$.
\end{proof}

\section{A minimal sufficient statistic can carry non-Halting information}

Proposition \ref{prop:MSSdiff} shows that the minimal sufficient statistic can carry a 
substantial amount of information that is not Halting information.

\begin{proposition} \label{prop:MSSdiff}
For $Z = S,P,F$:
$$
 \forall c\exists^{\infty} x\big[l^Z_x \geq^+ (k_x)^c \wedge I(x;H) \leq^+ k_x \big]. 
$$
$$
 \exists \nu>0 \exists^{\infty} x\big[l^Z_x \geq^+ \nu l(x) + k_x \wedge
I(x;H) \leq^+ k_x \big]. 
$$
\end{proposition}

First a sketch of the proof of the Proposition is given.
Let $x^*$ be a program of length $K(x)$ that produces $x$.
If $Z$ is a SS, than it will be shown that 
$$
 x^* \computes Z, K(Z).
$$
This means that a shortest program for $x$ generates $K(Z)$. 
If $Z$ where equivalent with $\Omega^{n,i}$ for some $i$, 
than $i$ can be computed from $x, K(x)$. However, an $x$ will be 
constructed such that $x^*$ has a computational $m$-depth of $i$, but
$i$ has a high complexity given $x^*$. This shows that $x^*$ does not 
compute $i$, and that there can be no
SS $Z$ of length $i$. Since $i$ has large complexity given $x^*$, 
also numbers close to $i$ have large complexity given $x^*$.
This will allow to derive lower bounds for the minimal sufficient
statistic relative to the $m$-depth.
Before the proof Proposition \ref{prop:MSSdiff} is given, 
Lemmas \ref{lem:lowerY}-\ref{lem:enoughIpart2} are proved.


%
%
%

\begin{lemma} \label{lem:lowerY}
 Let $x \in 2^n$, and $i\leq n/2$ such that  
$$
 \begin{array}{rll}
  K(x|i^*) &=^+& n \\
  x_i &=& 1.
 \end{array}
$$
There is an $y \in 2^{n/2}$ such that:
$$
 \begin{array}{rll}
  x^i &=& y^i \\
  y   &<& x^{n/2}   \\
  K(y) &=^+& n/2 \\
  K(i|y) &=^+& K(i) \\
  I(y;H) &\leq^+& i \\ 
 \end{array}
$$
\end{lemma}

\begin{proof}
 Applying additivity of prefix-free Kolmogorov complexity, equation
 \eqref{eq:Kadditivity}:
 \begin{eqnarray*}
  K(x^i|i^*) &=^+& K(x|i^*) - K(x_{i...n}|(x^i)^*, i^*) \\
	&\geq^+& n - (n - i) \\
	&\geq^+& i,
 \end{eqnarray*}
 and therefore: $K(x^i|i^*) =^+ i$.

 Choose $v \in 2^{n/2-i-1}$ such that 
 $$
  K^H(v|x^i, i^*) \geq n/2 - i - 1.
 $$
 Such $v$ always exists. Let $y=x^i0v$. Obviously, 
 the first two conditions of the Lemma are satisfied.

 \noindent
 Since $K(x^i) =^+ i$:
 $x^i \compeq (x^i)^*$.
 Applying additivity of prefix-free Kolmogorov complexity:
 \begin{eqnarray*}
  K(y|i^*) &=^+& K(x^i|i^*) + K(v|x^i, i^*) \\
		&=^+& i + n/2 - i - 1 \\
		&=^+& n/2.
 \end{eqnarray*}
 Therefore, also the third condition is satisfied.

 \noindent
 Remark that $y \compeq y^*$ such that:
$$
 \begin{array}{rll}
  K(i|y,n) &=^+& K(i,y) - K(y) \\
	   &=^+& K(y|i) + K(i) - K(y) \\
	   &\geq^+& n/2 + K(i) - n/2 \\
	   &=& K(i). 
 \end{array}
$$
 Therefore, also the forth condition is satisfied.

 \noindent
 Remark that: 
 \begin{eqnarray*}
  K^H(y) &\geq^+& K^H(v) \geq^+ n/2 - i \\
  K(y) &\leq^+& n/2. \\ 
 \end{eqnarray*}
 Therefore, also the fifth condition is satisfied.
\end{proof}

%
%
%

Lemmas \ref{lem:complexNeighbours} and \ref{lem:complexNeighbours2} show that if $i$
can not be computed from $x$, than also numbers in some neighbourhood
can not be computed from $x$.
Let $\log^{(k)} i$ be the $k$-th iteration $\log ... \log i$.

\begin{lemma} \label{lem:complexNeighbours}
 Let $c$ be constant, if
 $$
  K(i|x) \geq^+ \log i + \log^{(2)} i + \log^{(3)} i,
 $$
 than
 $$
  \min \{K(j|x) : i^{1/c} \leq j \leq i^c\} \geq \log^{(3)} i - O(\log^{(4)} i).
 $$ 
\end{lemma}

\begin{proof}
The proof of the conditioned version on $x$ is the same as the unconditioned version, 
which will be shown here.
 \begin{eqnarray}
  K(i) &=^+& K(i, \log i, \log^{(2)} i, \log^{(3)} i) \nonumber  \\
         &=^+& K(i| (\log i)^*, (\log^{(2)} i)^*, (\log^{(3)} i)^*) \nonumber  \\
         &   & + K(\log i| (\log^{(2)} i)^*, (\log^{(3)} i)^*) \nonumber  \\
         &   & + K(\log^{(2)} i| (\log^{(3)} i)^*) \nonumber  \\
         &   & + K(\log^{(3)} i). \label{eq:KiDecomp} 
 \end{eqnarray}
 Since $K(w|\log w) \leq^+ \log w$ and $K(w) \leq^+ 2\log w$, we have that
 \begin{eqnarray*}
  K(\log^{(2)} i) &\geq^+ & K(\log^{(2)} i |(\log^{(3)} i)^*) \\
		     &=^+ & K(i) -  K(i| (\log i)^*, (\log^{(2)} i)^*, (\log^{(3)} i)^*) \\
         	         && - K(\log i| (\log^{(2)} i)^*, (\log^{(3)} i)^*) - K(\log^{(3)} i). \\
		     &\geq^+& \log^{(3)} i - O(\log^{(4)} i).
 \end{eqnarray*}
 Remark that:
 $$
  \log^{(2)} i =^+ \log (1/c \log i) \leq \log^{2} j \leq  \log (c \log i) =^+ \log^{(2)} i. 
 $$
 therefore,
 $$ 
  K(j) \geq^+ K(\log^{(2)} j) \geq^+ \log^{(3)} i.
 $$
\end{proof}

\begin{lemma} \label{lem:complexNeighbours2}
 For any $c$, let $\tilde{i}$ be the $c$ most significant bits of $i$.
 If $i(1 - 2^{-c}) \leq j \leq i(1+2^{-c})$, than $K(j|n) \geq^+ K(\tilde{i}|n)$. 
\end{lemma}

\begin{proof}
Trivial. 
\end{proof}

In the proof of Proposition \ref{prop:MSSdiff} an $i$ will be needed that 
both satisfies the conditions of Lemmas \ref{lem:lowerY} and \ref{lem:complexNeighbours}. 
Lemmas \ref{lem:enoughI}, \ref{lem:enoughIpart1} and
\ref{lem:enoughIpart2} show that such $i$ can be constructed.

\begin{lemma} \label{lem:enoughI}
 For any $x$ with $K(x) \geq^+ n$, there is at least one $i$ satisfying 
 both conditions of Lemmas \ref{lem:lowerY} and \ref{lem:complexNeighbours}, 
 and there is at least one $i$ satisfying
 both conditions of Lemmas \ref{lem:lowerY} and \ref{lem:complexNeighbours}. 
\end{lemma}

\begin{proof}
 The first claim of the proposition implies the second claim, which is
 shown here.
 Remark that since $K$ is implicitly conditioned on $n$, it follows for 
 $\log i \geq (\log n)/2$ and $i\leq n$ that
 \begin{eqnarray*}
  K(i) &\leq^+& K(i|\log i) \\
	&\leq^+& \log i + 2\log (\log n - \log i) \\
	&\leq^+& \log n 
 \end{eqnarray*}
 This shows that for every $i \leq n$, there is a $p \in 2^{<\log n + O(1)}$, 
 such that $\Phi(p|n) \da= i$.
 Therefore, if $K(x) \geq^+ n$, than by Lemma 
 \ref{lem:enoughIpart1} for any $\nu$ there are maximally
 $\nu n$ different $i$ such that $K(x|i^*) \leq^+ n$.
 By Lemma \ref{lem:enoughIpart2} there are also only $n/8$ 
 different $i \leq n/2$ such that the condition of 
 Lemma \ref{lem:complexNeighbours} is not satisfied.
 Finally, there are maximally $n/4 + 2\log n + O(1)$ different 
 $i$ such that $x_i = 0$, since otherwise $x$ could be compressed.
 This shows that there are maximally 
 $$
  \nu n + n/4 + 2\log n + n/8 + O(1)
 $$
 many $i\leq n/2$ that not satisfy the conditions of Lemmas  
 \ref{lem:lowerY} and \ref{lem:complexNeighbours}. 
 Therefore, for $\nu$ sufficiently small, there must be at least
 one $i$ satisfying the conditions of 
 Lemmas \ref{lem:lowerY} and \ref{lem:complexNeighbours}. 
\end{proof}

\begin{lemma} \label{lem:enoughIpart1}
 Let $\nu > 0$, and
 $$ 
  S_{x,c} = \{p \in 2^{<\log n + O(1)}: K(x|p) \leq n-c \}.
 $$
 There is a $c$ such that for any $x$ with $K(x) \geq^+ n$:
 $$
  |S_{x,c} | \leq \nu n.
 $$
\end{lemma}

\begin{proof}
 Let 
 $$
  U_{x,c} = \{(p,q): q \in 2^{<n-c} \wedge p \in 2^{<n+O(1)} \wedge \Phi(q|p) \da = x \}.
 $$
 Suppose that 
\begin{equation} \label{eq:generalC}
  \forall c \exists x \big[ \, |S_{x,c} | \geq \nu n\big],
\end{equation} 
 than,
 $$
  \forall c \exists x \big[ \, |U_{x,c} | \geq \nu n\big],
 $$
 Let 
 $$
  P(x) = |U_{x,c}|n 2^{-n}.
 $$
 Remark that $U_{x,c}$ is enumerable, and therefore $P(x)$ can be 
 enumerated from $x,c$. 
 Applying the coding theorem shows that:
 \begin{eqnarray*}
  K(x) &\leq^+& -\log P(x) + K(P) \\
	&\leq^+& \log \nu + n - c + 2\log c. 
 \end{eqnarray*}
 Since $K(x) \geq^+ n$, this shows that
 $$
  c - 2\log c \leq^+ \nu.
 $$
 Which contradicts the generality of $c$ from equation
 \eqref{eq:generalC}.
\end{proof}

\begin{lemma} \label{lem:enoughIpart2}
 There are $\frac{3n}{8}$ many $i < \frac{n}{2}$ satisfying the 
 condition of Lemma \ref{lem:complexNeighbours}.  
\end{lemma}

\begin{proof}
 Let $\log^{(0)} i = i$. 
 Let $0 \leq j \leq c = c' -1 = 3$, 
 there are maximally $n2^{-c'-1}$ many $i < n/2$ that not satisfy:
 $$
  K(\log^{(j)} i | (\log^{(j+1)} i)^*, ...,  (\log^{(c)} i)^*, n) \geq \log^{(j+1)} i - c'.
 $$ 
 Therefore, maximally $(c+1)2^{-c'-1}n=\frac{n}{8}$ many  
 $i < \frac{n}{2}$ do not 
 satisfy the above equation for some $j=0,...,c$.
 The decomposition in equation \eqref{eq:KiDecomp} 
 finishes the proof.
\end{proof}

\textit{Proof of proposition \ref{prop:MSSdiff}}
Let $m_t$ be an enumeration of the universal enumerable semimeasure $m$, such that 
that for all $t$ there is maximally one $x \in 2^n$ with
$$
 m_t(x) \not= m_{t+1}(x).
$$
Additionally assume that for all $k < 2^{n/2}$, for witch there is a $t$ such that
\begin{equation} \label{eq:timeStep}
 \sum \{ m_{t-1}(x): x \in 2^n \} < k2^{-n/2} \leq \sum \{m_t(x): x\in 2^n\},
\end{equation} 
there is a $z_k \in 2^n$, such that $m_t(z_k) \leq 2^{-n}$ and $m_{t+1}(z_k) > 2^{-n}$.
Remark that for any such $k$ 
$$
 z_k \compeq k.
$$ 
Remark that equation \eqref{eq:timeStep} is very similar to the
requirement $\Omega^n_t < k2^{n/2} \leq \Omega^n_t$. However, 
to reduce technical details, this equivalent formulation of the
proof was preferred.

By Lemma \ref{lem:complexityOmega} one has $K(\Omega^{n,n}|n) \geq^+ n$. 
For $n$ large enough, let $y \in 2^{n/2}$ as in Lemma \ref{lem:lowerY} 
with $x = \Omega^{n,n}$, and $i$ chosen such that 
\begin{equation} \label{eq:requirementsI}
 \begin{array}{rll}
  x_i &=& 1  \\
  K(x) &=^+& n,
 \end{array}
\end{equation}
and $K(i|x)$ large enough such that it will satisfy some upper bounds
determined later in the proof. 
Remark that $x_i = \Omega^n_i = 1$ and $y_i=0$, and therefore
$\Omega^{n,i-1} \leq y < \Omega^{n,i}$. This shows that $y$ determines a
$k \leq 2^{n/2}$ such that equation \eqref{eq:timeStep} is satisfied, and
since $\Omega^{n,i-1} \leq \sum\{m_t(x):x \in 2^n\}$, the corresponding $t$ 
satisfies $t \geq t_{i-1}$.
Let $z=z_{k}$.
Remark that 
$$
 z \compeq y.
$$
This implies that $K(z) \leq^+ n/2$.   , therefore $i \leq^+ k_z$. 
At time $t_i$, one has $m_t(z) \geq 2^{-n}$, and by 
a time-bound version of the coding theorem, $K_{t_{i-O(1)}}(z) \geq^+ n$. 
By Lemma \ref{lem:lowerY}, $I(y;H) \leq^+ i \leq^+ k_z$. Therefore, 
$z$ satisfies the right condition of both claims of Proposition \ref{prop:MSSdiff}.

Let $Z$ be a minimal SS.
xhx
Since $w$ can be computed by first computing $Z$,
and than the corresponding information of $||\log Z||$, and the total code to do this is 
shorter than $K(z) + O(1)$, it follows by Lemma
\ref{lem:generateWithnesses} that:
\begin{equation} \label{eq:zStarComputesLSz}
 z^* \computes Z \computes l^Z_z.
\end{equation} 

Now the left condition of the first claim of Proposition \ref{prop:MSSdiff} is shown.
Choose in addition of the requirements mentioned in \eqref{eq:requirementsI} the 
upper bound for $K(i)$ of Proposition \ref{lem:complexNeighbours}.
$$
 K(i|z) =^+ K(i) \geq^+ \log i + \log^{(2)} i + \log^{(3)} i.
$$
Such $i$ exists by Lemma \ref{lem:enoughI}. Lemma \ref{lem:complexNeighbours} shows 
that for any $j$ with $i^{1/c} \geq j \geq i^c$: 
$$
 K(j|z) \geq \log^{(3)} i - O(\log^{(4)}),
$$
and therefore, assuming $\log^{(3)} i > O(1)$ one has:
\begin{equation} \label{eq:zNotComputesJ}
 z^* \not\computes j.
\end{equation} 
Combined with equation \eqref{eq:zStarComputesLSz}, this shows that 
either $l^Z_z < i^{1/c}$ or either $l^Z_z > i^c$.
By Proposition \ref{prop:MSSvsMdepth} it follows that 
$$ 
 l^Z_x \geq^+ k_x - 2\log k_x \geq^+ i - 2\log i,
$$
and therefore, $l^Z_z > i^c$. This shows the left condition of the first claim of Proposition \ref{prop:MSSdiff}.

Now the left condition of the second claim  is shown.
Let $c > O(1)$, and choose for some $2^{c-1} \leq \tilde{i} \leq 2^c$,
such that $K(i|x) \geq c$.
Let $i = \tilde{i}2^{\log n-c-1}$.  Remark that $i = O(n)$. 
By Lemma \ref{lem:complexNeighbours2}, 
for $i(1-2^{-c}) \leq j \leq i(1+2^{-c})$ we have $K(j|x) \geq^+ c$ and therefore 
equation \eqref{eq:zNotComputesJ} holds. The same reasoning as in the previous 
paragraph shows the left condition of the second claim of Proposition \ref{prop:MSSdiff}.
\qed

Proposition \ref{prop:MSSdiff} shows that there can be a 
difference between the minimal SS and 
the information carried in the initial bits of the Halting sequence.
However, the proposition does not address the question whether this
difference is substantial with respect to an attempt to interpret algorithms 
that were designed inspired by the use of minimal SS. 
The first claim of Proposition \ref{prop:MSSdiff}
can only be satisfied for $n$ sufficiently large, compared to the $O(1)$ constants.
To obtain equation \eqref{eq:zNotComputesJ} it is assumed that
$\log^{(3)} i \geq O(1)$, therefore,
$$
 n>i> 2^{2^{2^{O(1)}}}.
$$
Even if it is assumed that the arbitrary constants are very low, suppose 
that $O(1)=4$ could be chosen in the above equation, the corresponding $n$ 
is much larger than the length of any data that can possibly be the 
input of an algorithm.  In the proof of the second equation of 
Proposition \ref{prop:MSSdiff},
the constructed $\nu$ satisfies $\nu \leq 2^{-c}$, which implies that for 
large $c$ the largest fraction of the information of the minimal SS of 
the constructed $z$ in the proof is Halting information.
Therefore the result in this paper is only a partial result addressing 
the possible interpretation of the minimal SS as 
containing Halting information.

\section{Weak sufficient statistics} \label{sec:WSS}

A variant of the definition of a SS is proposed: the weak sufficient statistic (WSS). 
A criterion is provided for which the WSS is 
It is defined such that the minimal WSS is equivalent with
an initial segment of the Halting sequence relative to a plain Turing machine. 
An explicit construction will be given to convert an initial 
segment of the Halting sequence into a minimal WSS and to convert a 
minimal WSS into an initial segment of the Halting sequence.

The reason why a minimal SS, as defined higher is not equivalent with an 
initial segment of the Halting sequence, is that the length of that 
segment carries information that would be available in the description 
of $x$, while this information does not contribute to the compression of $x$. 
If the minimal SS is encoded such that the information of the length of 
the minimal SS does not ``count", than there is an equivalence. It turns out that
this is possible by conditioning the complexities of $x,Z$ on $C(Z)$ 
in the definition of a SS, where $C(Z)$ is the Kolmogorov
complexity with respect to a plain Turing machine. 
Let $\Psi$ a plain Turing Machine, than $C(x) = \min\{l(p): \Psi(p,n) \da= x\}$.
The following equation relates prefix-free and plain Kolmogorov
complexity \cite{LiVitanyi}: 
\begin{equation} \label{eq:CvsK}
 C(x) =^+ K(x|C(x))
\end{equation}

\begin{definition} \label{def:WSS}
Let $x \in 2^n$. 
\begin{itemize} 
 \item A finite set $S \subset 2^n$ is a {\em weak sufficient set statistic} of a binary string $x$ iff $x\in S$ and
  \begin{equation} \label{eq:setWSS}
   C(S) + \log |S| =^+ K(x|C(S)).
  \end{equation} 
 \item A computable semimeasure $P$ over $2^n$ is a {\em weak sufficient probabilistic statistic} of a binary string $x$ iff 
  \begin{equation} \label{eq:probWSS}
   C(P) - \log P(x) =^+ K(x|C(P)).
  \end{equation} 
 \item A total function $F:2^{<n} \rar 2^n$ is a {\em weak sufficient function statistic} of a binary string $x$ iff 
  $$
   C(F) - \log P(x) =^+ K(x|C(F)).
  $$
\end{itemize}
For $Z=S,P$, the {\em minimal weak sufficient statistic} $Z'_x$ is the
weak sufficient statistic $Z$ such that $C(Z)$ is minimal
within some constant.
Let $l'^Z_x = C(Z'_x)$.
\end{definition}

In the same way as in Lemma \ref{prop:SSmodels}, for any $x$, a probabilistic
weak sufficient statistic (probabilistic WSS) is algorithmically
equivalent with an function WSS.

Let $||\log Z||$ be either $\log |S|, -\log P(x)$, or $\min \{l(d):
F(d) = x$. Then de defining equation for a WSS is given by:
$$
 C(Z) + ||\log Z|| =^+ K(x|C(Z)).
$$
By Lemma \ref{lem:generateWithnesses}, 
it follows that there are only 
a finite amount of SS'es. 
By Proposition \ref{prop:manyWSSes} there 
can be an large amount of WSS'es for a string $x$.

\begin{proposition} \label{prop:manyWSSes} 
 If $K(x) \geq^+ n$, than $x$ has $O(n)$ different WSS'es.
\end{proposition}

\begin{proof}
 Let $i$ such that $K(x|i) =^+ n$. By Lemma \ref{lem:enoughIpart1} 
 there are $O(n)$ such $i$.
 In the same way as the beginning of the proof of Lemma
 \ref{lem:lowerY}, it follows that
  $K(x^i|i) =^+ i$.
 Let 
 $$
  S_i = \{x^iv: v \in 2^{n-i}\}. 
 $$
 Remark that $K(S_i|i) =^+ K(x^i|i) =^+ i$ and thus by equation 
 \eqref{eq:CvsK} $C(S_i) =^+ i$. Also remark that $\log |S_i| = n-i$.
 This shows that $S_i$ satisfies equation \eqref{eq:setWSS}.
\end{proof}


\begin{proposition} \label{prop:WSSsuperset}
 For $Z = S,P,F$, if $Z$ is a SS of $x \in 2^n$, and 
 $$
  Z,K(Z) \computes C(Z),
 $$
 than $Z$ is a WSS of $x$. 
\end{proposition}

\begin{proof}
 Remark that by equation \eqref{eq:CvsK} every WSS $Z$ defines a shortest description of 
 $x$ given $C(Z)$ on a prefix-free Turing machine.
 By the conditioned version of Lemma \ref{lem:generateWithnesses}, 
 it follows that 
 $$
  x,K(x|C(Z)), C(Z) \computes Z.
 $$
 By the assumption of the proposition
 $$
  Z^* \rar C(Z).
 $$
 One also has $K(x) = K(x,K(x))$, and its conditioned equivalent.  Therefore:
 \begin{eqnarray*}
  K(x|C(Z)) & =^+ & K(x, K(x|C(Z)) |C(Z)) \\
  	& =^+ & K(x, Z |C(Z)) \\
	& =^+ & K(x | Z^*, C(Z)) + K(Z|C(Z)) \\
	& =^+ & K(x | Z^*) + K(Z|C(Z)) \\
	& =^+ & K(x) - K(Z) + K(Z|C(Z)) 
 \end{eqnarray*}
 $$
  ||\log Z|| =^+ K(x) - K(Z) =^+ K(x|C(Z)) - C(Z). 
 $$
\end{proof}

The question raises whether $Z,K(Z) \computes C(Z)$. Let $^k2$ be the tetration with 
base 2 and height $k$, it is the $k$-th iteration of taking the power of 2, it is:
$$
 2^{(2^{(...^2)})}.
$$
The inverse of the tetration function is the super-logarithm, it is 
$$
 \slog x = \max \{k: ^k2 \leq x\}.
$$

\begin{lemma} \label{lem:tetration}
 $$
  K(C(x)|x,K(x)) \leq^+ O(\slog x).
 $$ 
\end{lemma}

\begin{proof}
 $C(x)$ is approximated as:
 $$
  \begin{array}{rclcl}
  k_1 & = & K(x) \\
  k_2 & = & K(x|k_1^*) & = & K(x|K(x)^*) \\
  k_3 & = & K(x|k_2^*) & = & K(x|K(x|K(x)^*)^*) \\ 
  k_i & = & K(x|k_{i-1}^*) & = & K(x|K(x|... ^*)^*).
  \end{array}
 $$
 Remark that since $k_1 \leq^+ 2 \log x$, it follows that 
 $k_1 - k_2 \leq^+ 2\log^{(2)} x$. 
 Suppose that 
 $$
  \abs(k_{i-1} - k_{i}) \leq^+ 2\log^{(i)} x,
 $$ 
 than it follows that
 \begin{eqnarray*}
  \abs(k_{i} - k_{i+1}) &\leq^+& \abs( K(x|k_i) - K(x|k_{i+1})) \\
		&\leq^+& 2 \log \abs (k_i - k_{i+1}) \\
		&\leq^+& 2\log^{(i+1)} x.
 \end{eqnarray*}
 and therefore the series has converged after $\slog x$ steps, 
 within a constant. 
 The limit of the series is some $k$ for which $K(x|k) =^+ k$. 
 There is only one value $k$ that for some $x$ satisfies $K(x|k) =^+ k$.  
 Since if there was also a $l<k$ such that $K(x|l) =^+ l$, than 
 $$
  k - l =^+ K(x|k) - K(x|l) \leq^+ 2 \log(k - l), 
 $$
 and therefore, $k =^+ l$. Remark that the proof of equation
 \eqref{eq:CvsK}, see \cite[Lemma 3.1.1]{LiVitanyi} 
 also shows that  
 $$
  C(x) =^+ K(x|C(x)^*).
 $$
 Therefore, it follows that this series $k_i$ converges to $C(x)$. 
 To prove the proposition, it suffices to show that the evaluation 
 of $k_{i+1}, K(x,k_{i+1})$ from 
 $k_i,K(x,k_i),x$ requires at most a constant amount of bits.
 First remark that for any $u,v$ \cite{decomposition}:
 $$
  K(u,v) =^+ K(K(u|v^*),u,v).
 $$
 Since there are maximally a constant amount of programs of length $K(u,v)$, that produce 
 $u,v$, $K(u|v^*)$ can be found within $O(1)$ bits from $u,v,K(u,v)$. 
 Replacing $u = x$ and $v = k_i$, shows that $k_{i+1}$ can be computed from 
 $x,k_i,K(x,k_i)$. In a similar way, it is shown that $K(x,k_{i+1})$ can be computed from
 $K(x,k_i,k_{i+1})$. Therefore, $k_{i+1}, K(x,k_{i+1})$ can be computed from 
 $k_i, K(x,k_i)$.
\end{proof}

By Lemma \ref{lem:tetration} and Proposition \ref{prop:WSSsuperset}, it can be stated
that for strings of realistic length, every WSS is a SS. This is why the name 
\textit{weak} sufficient statistic was chosen. It contrasts with the
name \textit{strong} sufficient statistic defined in \cite{SSS}.

An explicit construction of a probabilistic WSS $P'_x$ for an $x \in
2^n$ is now given. Remark that 
in \cite{algorithmicStatistics} a construction is given of what is called an 
``Explicit minimal near-sufficient statistic". The construction there 
can be adapted to a construction of a set WSS using the same ideas as
as the construction of $P'_x$. 
The construction of $P'_x$ 
makes use of $k'_x$, a variation of 
$m$-depth, which will be called $BB$-depth since it uses the 
 Buzzy Beaver function. Assume $c$ be large enough:
\begin{eqnarray*}
 BB(k) &=& \max \{\Psi(p): p \in 2^k\} \\ 
 k'_x &=& \min \{k: K_{BB(k)}(x|k) =^+ K(x|k)\} \\
 P'_x(y) &=& \begin{cases}  2^{-K_{BB(k'_x)}(y|k'_x) + k'_x - c} & \text{ if $ K_{BB(k'_x)}(y|k'_x) \not=^+ K_{BB(k'_x-1)}(y|k'_x)$,} \\
				0 				& \text{ otherwise.} \end{cases}. 
\end{eqnarray*}

\begin{proposition} \label{prop:PxIsPWSS}
 $P'_x$ is a probabilistic WSS.
\end{proposition}

Let $k_{x|l}$ be the conditional $m$-depth, it is the depth according
definition \ref{def:depth} with the 
semimeasure $m$ replaced by the conditional semimeasure $m(.|l)$. 
Then the following Lemma shows a relation between conditional $m$-depth
and $BB$-depth, which is similar to equation \eqref{eq:CvsK}.

\begin{lemma} \label{lem:mdepthVsBBdepth}
 $$
  k'_x =^+ k_{x|k'_x}
 $$ 
\end{lemma}

\begin{proof}
It suffices to show that:
\begin{eqnarray*}
 BB(k) &\leq& t_{k+O(1)|k} \\ 
 t_{k|k} &\leq& BB(k+O(1)) 
\end{eqnarray*}
The first inequality follows by remarking that any program of length $k$ halting on a 
plain Turing machine,  
can be adapted to a program of length $k+O(1)$ by adding a constant
amount of instructions, halting on a prefix-free Turing machine given $k$.

The second inequality follows by remarking that $\Omega^{n,k}_{|k}$, the conditional version of 
$\Omega^{n,k}$, defines a Halting program on plain Turing machine that outputs $t_{k|k}$ by 
adding a finite amount of instructions.
\end{proof}

\textit{Proof of Proposition \ref{prop:PxIsPWSS}}.
 First it will be shown that $P'_x$ is a semimeasure.  Let 
\begin{eqnarray*} 
     m_t(y|l) & = & 2^{-K_t(y|l)} 
\end{eqnarray*} 
 For $c'$ large enough, by Lemma \ref{lem:mdepthVsBBdepth}:
\begin{eqnarray*} \label{eq:someLB}
  && \sum_y m_{BB(k)}(y|k) - m_{BB(k-1)}(y|k) \\ 
  && \leq \sum_y m_{t_{k+c'|k}}(y|k) - m_{t_{k-c'|k}}(y|k) \\
  && \leq \Omega_{|k}^n - \Omega^n_{|k,t_{k-c'|k}} \leq 2^{-k+c'}.
\end{eqnarray*} 
 Choosing $c$ in the definition of $P'_x$ large enough, shows that $P'_x$
 is a semimeasure.

 Now it remains to show that $P'_x$ satisfies the defining equation 
 \eqref{eq:probWSS} of a probabilistic WSS. 
 Remark that given $C(P)$, a program for $P$ on a plain Turing machine
 can be turned into a program for $P$ given $C(P)$ on a prefix-free Turing machine 
 by adding a constant amount of instructions. Using Shannon-Fano code 
 \cite{LiVitanyi}, this shows that $C(P) - \log P(x) \geq^+ K(x|C(P))$.
 By the choice of $m$, one also has that 
 $$
  P'_x(x) = m_{BB(k)}(x|k) - m_{BB(k-1)}(x|k) \geq 1/2 m_{BB(k)}(x|k) =
2^{-K(x) + k'_x - c - 1}.
 $$
 This shows the $\leq^+$ inequality of equation \eqref{eq:probWSS}. 
\qed

\begin{proposition} \label{prop:WSSequivalence}
 $$
  P'_x, k'_x \compeq BB(k'_x), k'_x
 $$
\end{proposition}

Before the proof is given, first some technical result is written out:

\begin{lemma} \label{lem:BBcomputesBB}
For any $t \geq BB(k-O(1))$:
$$
 t,k \computes BB(k).
$$ 
\end{lemma}

\begin{proof}
 Let $c$, such that $t \geq BB(k-c)$.  
 Remark that $t,k \computes BB(k-c)$. Let 
 $$
  S_k = \{p \in 2^k: \Psi(p) \geq BB(k-c) \}.
 $$
 Suppose that $|S_k| > f(k)$, with $f$ unbounded.
 Let $p_1,p_2,...$ an enumeration of all binary strings in $2^k$,
 ordered with increasing computation time on $\Psi$.
 Remark that $l \compeq p_l$.
 Given $BB(k-c),k$ the set $S_k$ appears at the end of this
 enumeration. Therefore, there is some element $p_l \in S_k$,
 such that its index $l$ ends with $\log f(k)-1$ zeros.
 $l$ has plain complexity below $k - \log f(k) + 2\log \log f(k)$.
 Therefore, $l$ can be transformed into a program that 
 has an output above $BB(k-c)$, and has length unboundedly 
 below $k-c$, which contradicts the definition of $BB(k-c)$.
\end{proof}

\textit{Proof of Proposition \ref{prop:WSSequivalence}}
The left $\longleftarrow$ follows from the definition of $P'_x$. It remains to show the 
right $\computes$. By Lemma \ref{lem:BBcomputesBB}, it suffices to show that 
$P'_x,k'_x \computes t,k'_x$ with $t \geq BB(k'_x-O(1))$.
Let $z$ be the lexicographic first string with $m_{BB(k'_x-c)}(y|k'_x)
\geq 2^{-n}$, for some 
constant $c$ large enough, than it follows that 
$$
 m_{BB(k'_x)}(y|k'_x) \leq^+ k'_x + 2\log k'_x.
$$
Therefore, by estimating $BB(k'_x),BB(k'_x-1)$ on $\Psi_t$ for
increasing $t$, and using $k'_x$, one can only find 
an equality for $P'_x(z)$, for $t \geq BB(k'_x-c)$. 
Therefore, $P'_x,k'_x \computes t,k'_x$.
\qed

\section{Minimal typical model}


Typical set models were studied in \cite{structureFunction}, and it was
shown that within logarithmic bounds,
 the complexity of the minimal typical set and the minimal SS where equal. 
It is shown here that a minimal typical model is 
equivalent with a minimal WSS and therefore, there complexities are
equal within constant bounds. For either set, 
probabilistic and functional models. Therefore, 
the minimal typical model is also equivalent to some initial segment of
the Halting sequence.

\begin{definition} \label{def:minSufStat}
\begin{itemize} 
 \item Let $S^*$ denote the shortest program on a plain Turing machine.
  A finite set $S$ is a {\em typical set} for a binary string $x$ iff $x\in S$ and
  $$
   \log |S| =^+ K(x|S^*).
  $$
 \item Let $P^*$ denote the shortest program on a plain Turing machine
that computes $P$.
A computable semimeasure $P$ is a {\em typical semimeasure} for a binary string $x$ iff 
  $$
   - \log P(x) =^+ K(x|P^*).
  $$
 \item Let $F^*$ denote the shortest program on a plain Turing machine
that computes $F$.
  A computable function $F:\omega \rar \omega$ is a {\em typical function} for a binary string $x$  iff
  $$ 
   \exists d[F(d) = x  \wedge l(d) =^+ K(x|F^*).  
  $$
\end{itemize}
For $Z=S,P,F$, a {\em minimal typical model} is a typical model $Z$ such that $K(Z)$ is minimal within a constant.
\end{definition}

The same proof of Proposition \ref{prop:SSmodels} also shows that the set of function typical
models is the same as probabilistic typical models.
Remark that in \cite{structureFunction}, a set typical model is defined as
$\log |S| =^+ K(x|S)$. In this definition $S$ is replaced by its minimal
description, with respect to a plain Turing machine. 
Since \cite{structureFunction} only considers equalities of functions
within logarithmic terms of $n$ both in value and in argument.  
The results shown there, also remain valid using the
definition above.
By Lemma \ref{lem:tetration}, the results also hold within $O(\slog)$ terms,
if $Z^*$ was the shortest representation on a prefix-free Turing
machine.


\begin{proposition} \label{prop:WSSisTM}
Every WSS for $x \in 2^n$ is also a typical model (TM) for $x \in 2^n$.
\end{proposition}

\begin{proof}
 Remind that for any WSS $Z$:
 $$
  x,C(Z),K(x|C(Z)) \computes Z.
 $$
 Therefore:
 \begin{eqnarray*}
  K(x|Z^*) &=^+& K(x|Z^*, C(Z)) \\
	&=^+& K(x,Z|C(Z)) - K(Z|C(Z)) \\
	&=^+& K(x|C(Z)) - K(Z|C(Z)) \\
	&=^+& ||\log Z||, 
 \end{eqnarray*}
 where $||\log Z||$ is either $\log |S|, -\log P(x)$, or $\min \{l(d):
F(d) = x \}$. 
\end{proof}

By the same example as in \cite{structureFunction}, 
it follows that there are TM's that are not WSS'es.
According to Proposition \ref{prop:minimalTM}, $P'_x$ defines
a minimal TM and by Corollary \ref{cor:TMvsHalting} 
a minimal WSS is equivalent with the 
minimal TM, which is equivalent with an initial segment of the Halting sequence.

\begin{proposition} \label{prop:minimalTM}
 If $P$ is a probabilistic TM for $x$, than $C(P) \geq^+ k'_x$.
\end{proposition}

Before the proposition is proved, Lemma \ref{lem:timeBoundAdditivity} is shown.

\begin{lemma} \label{lem:timeBoundAdditivity}
 For some large enough computable function $f$:
 $$
  K_t(x,y) \geq^+ K_{f(t,n)}(x) + K_{f(t,n)}(y|x, K_{f(t,n)}).
 $$ 
\end{lemma}

This proof is essentially the same as additivity of prefix-free
Kolmogorov complexity \cite{LiVitanyi}, but formulated with time-bounds. 

\begin{proof}
 Let 
 \begin{eqnarray*}
  m_t(x,y)&=& \sum \{2^{-l(p)}: \Phi_t(p) \da=[x,y] \} \\
  S_x &=& \{p: \Phi_t(p) \da=[x,z] \wedge z \in 2^n\}.
 \end{eqnarray*}
 Remark that $S_x$ can be enumerated from $x$, and by the coding theorem:
 $$
  K_{f(t,n)}(x) \leq^+ -\log \sum_z m_t(x,z) = -\log \sum \{2^{-l(p)} : p \in S_x \}.
 $$
 Therefore, 
 $$
  P(z) = 2^{K_{f(t,n)}(x) - O(1)} m_t(x,z)
 $$ 
 defines a conditional semimeasure that can be computed from $x,K_{f(t,n)}(x)$ in 
 time $t$. Shannon Fano code shows that for $f$ large enough:  
 $$
  K_{f(t,n)}(y|x, K_{f(t,n)}(x)) \leq^+ K_t(x,y) - K_{f(t,n)}(x).
 $$ 
\end{proof}

\textit{Proof of Proposition \ref{prop:minimalTM}}
 Let $P$ be a TM, than it will be
shown that 
\begin{equation} \label{eq:xCompressible}
 K(x|C(P)) =^+ K_{BB(C(P)+O(1))}(x|C(P)).
\end{equation} 
If $C(P)$ was unboundedly below $k'_x$, this would contradict the definition of
$k'_x$. Therefore it remains to show equation \eqref{eq:xCompressible}. 
\begin{eqnarray}
 C(P) - \log P(x) &=^+& C(P) + K(x|P^*) \nonumber \\
		&=^+& K(P|C(P)) + K(x|P^*, C(P)) \\
		&=^+& K(x,P|C(P)) \nonumber \\
		&=^+& K(x|C(P)) + K(P|x,K(x|C(P)),C(P)).
			\label{eq:CPfirst}
\end{eqnarray}
On the other side, let $s$ be the computation time to compute 
$-\log P(z)$ for all $z \in 2^n$ from $P^*$. Than 
$K_s(x|P^*) =^+ - \log P(x)$. For computable functions $f,g$ 
large enough we have by Lemma \ref{lem:timeBoundAdditivity}:
\begin{eqnarray*}
 C(P) - \log P(x) &=^+& C(P) + K_s(x|P^*) \nonumber \\
		&\geq^+& K_{g(s)}(x,P|C(P)) \nonumber \\
		&\geq^+& K_{f(s)}(x|C(P)) +
	K_{f(s)}(P|x,K_{f(s)}(x|C(P)),C(P)).  \label{eq:CPsecond}
\end{eqnarray*}
Let $\Delta = K_{f(s)}(x|C(P)) - K(x|C(P)) \geq 0$, than combining equations
\eqref{eq:CPfirst} and \eqref{eq:CPsecond}:
\begin{eqnarray*}
 && K(x|C(P)) + K(P|x,K(x|C(P)),C(P)) \\
 && \geq^+ K_{f(s)}(x|C(P)) + K_{f(s)}(P|x,K_{f(s)}(x|C(P)),C(P)) \\ 
 && \geq^+ K(x|C(P)) + \Delta + K(P|x,K(x|C(P)),C(P)) - 2\log \Delta.
\end{eqnarray*}
This shows that $0 \geq^+ \Delta - 2\log \Delta$, and therefore 
$ \Delta =^+ 0$. Since $BB(C(P) + O(1)) \geq s$, equation
\eqref{eq:xCompressible} is satisfied.
\qed

\begin{corollary} \label{cor:PxIsTM}
 $P'_x$ defines a minimal typical probabilistic model.
\end{corollary}

\begin{proof}
 Since $C(P'_x)$ is a WSS, it is also a TM, and since
 $C(P'_x) =^+ k'_x$, there is no TM which is smaller
 by more than a constant. 
\end{proof}

Let $H'^n$ be the Halting sequence relative to a plain Turing
machine, conditioned. It is, $H'^n_i = 1$ if $\Psi(i,n) \da$,
and $H'^n_{i}=0$ otherwise. Corollary \ref{cor:TMvsHalting} shows 
that a probabilistic minimal TM is equivalent with an 
initial segment of $H'_{|l}$.

\begin{corollary} \label{cor:TMvsHalting}
 If $P$ is a minimal typical probabilistic model, and 
 $P^*$ its minimal description on a plain Turing machine, than 
 $$
  P^* \compeq  H'^{n,2^{k'_x}} \compeq (P'_x)^*.
 $$
\end{corollary}

\begin{proof}
 Remark that by Corollary \ref{cor:PxIsTM}, we have that $C(P) = k'_x$.
 From the proof of Proposition \ref{prop:minimalTM} equation \eqref{eq:xCompressible}  
 actually shows that if $s$ is the maximal to evaluate a Shannon-Fano
 code according to $P(y)$ for any $y \in 2^n$, than:
 $$
  K(x|C(P)) =^+ K_{s}(x|C(P)),
 $$
 This shows that $s \geq BB(k'_x-O(1))$.
 Remark that $s$ can be computed from $P$,
 therefore, 
 $$
  s \leq BB(C(P)+O(1)) \leq BB(k'_x + O(1)).
 $$
 Let $p$ be the program of length $k'_x$ with largest output, 
 than 
 $$
  P \compeq p \compeq H'^{n,2^{k'_x}}.
 $$ 
 The last $\compeq$ follows from Proposition \ref{prop:WSSequivalence}.
\end{proof}

\vspace{1cm}

\begin{acknowledgement}
The author is grateful to P. Vitanyi who raised the question on the relation between 
an algorithmic minimal sufficient statistics and an initial segment of
a Halting sequence. 
\end{acknowledgement}

\bibliographystyle{plain}
\bibliography{../bib/practCausalities,../bib/statisticalCausalities,../bib/eigen,../bib/bib,../bib/kolmogorov}

\begin{thebibliography}{10}

\bibitem{sophistication}
Luis Antunes and Lance Fortnow.
\newblock Sophistication revisited.
\newblock {\em Theor. Comp. Sys.}, 45(1):150--161, 2009.

\bibitem{Antunes}
Luis Antunes, Lance Fortnow, and Dieter~Van Melkebeek.
\newblock Computational depth.
\newblock {\em Computational Complexity, Annual IEEE Conference on}, 0:0266,
  2001.

\bibitem{decomposition}
B.~{Bauwens}.
\newblock {Additivity of on-line decision complexity is violated by a linear
  term in the length of a binary string}.
\newblock {\em ArXiv e-prints}, August 2009.

\bibitem{LuminyTalk}
Bruno Bauwens.
\newblock Ideal hypothesis testing and algorithmic information transfer, june
  2009.
\newblock Talk in \textit{Conference on Logic, computability and randomness},
  www.lif.univ-mrs.fr/lce/bauwens.pdf.

\bibitem{mdepth}
Bruno Bauwens.
\newblock \textit{m}-depth.
\newblock In preparation, 2009.

\bibitem{Cover}
T.M. Cover and T.A. Joy.
\newblock {\em Elements of Information Theory}.
\newblock John Wiley \& sons, 1991.

\bibitem{minSuffStatJPG}
S.~de~Rooij and P.M.B. Vit{\'a}nyi.
\newblock Approximating rate-distortion graphs of individual data: Experiments
  in lossy compression and denoising.
\newblock {\em CoRR}, abs/cs/0609121, 2006.

\bibitem{algorithmicStatistics}
P.~G{\'a}cs, J.~Tromp, and Vit{\'a}nyi P.M.B.
\newblock Algorithmic statistics.
\newblock {\em IEEE Transactions on Information Theory}, 47(6):2443--2463,
  2001.

\bibitem{complexityOfComplexity}
Peter Gacs.
\newblock On the symmetry of algorithmic information.
\newblock {\em Soviet Mathematical Dokledy}, 15:1477--1480, 1974.

\bibitem{GacsNotes}
Peter Gacs.
\newblock Lecture notes on descriptional complexity and randomness.
\newblock Unpublished, 2009.

\bibitem{LiVitanyi}
Ming Li and Paul~M.B. Vitanyi.
\newblock {\em An Introduction to Kolmogorov Complexity and Its Applications}.
\newblock Springer-Verlag, New York, 2009.

\bibitem{structureFunction}
Nikolai~K. Vereshchagin and Paul M.~B. Vit{\'a}nyi.
\newblock Kolmogorov's structure functions and model selection.
\newblock {\em IEEE Transactions on Information Theory}, 50(12):3265--3290,
  2004.

\bibitem{SSS}
Nikolai~K. Verschagin.
\newblock Agorithmic minimal sufficient statistics: a new definition.
\newblock Presented on the 4th conference on randomness, computability and
  logic, Luminy., nov 2009.

\bibitem{meaningfulInformation}
P.~M. Vitanyi.
\newblock Meaningful information meaningful information.
\newblock {\em Information Theory, IEEE Transactions on}, 52(10):4617--4626,
  2006.

\bibitem{idealInduction}
Paul M.~B. Vit{\'a}nyi and Ming Li.
\newblock Minimum description length induction, bayesianism, and kolmogorov
  complexity.
\newblock {\em IEEE Transactions on Information Theory}, 46(2):446--464, 2000.

\end{thebibliography}

\end{document}